\documentclass[11pt,letterpaper]{article}

\usepackage{mathpple}
\usepackage{graphicx}

\usepackage{geometry}
\begin{document}

\thispagestyle{empty}


{\small
\textbf{Preprint of:}\\
S. Bayoudh,
T. A. Nieminen,
N. R. Heckenberg and
H. Rubinsztein-Dunlop\\
``Orientation of biological cells using plane-polarized Gaussian beam
optical tweezers''\\
\textit{Journal of Modern Optics} \textbf{50}(10), 1581--1590 (2003)\\
Changes: Corrections to equations (6) and (7).
}

\hrulefill

\vspace{7mm}

\begin{center}

\Large {\bf
Orientation of biological cells using plane-polarized\\ Gaussian beam
optical tweezers \\
}

\vspace{1pc}
\large

S. Bayoudh,
T. A. Nieminen,
N. R. Heckenberg and
H. Rubinsztein-Dunlop

\vspace{1pc}
\small
\textit{
Centre for Biophotonics and Laser Science, Department of Physics \\
The University of Queensland, Brisbane, QLD 4072, Australia.\\
tel: +61-7-3365 3405, fax: +61-7-3365 1242, \\
e-mail: timo@physics.uq.edu.au
}
\vspace{1pc}

\end{center}

\begin{abstract}

Optical tweezers are widely used for the manipulation of cells and their
internal structures. However, the degree of manipulation possible is limited
by poor control over the orientation of trapped cells.
We show that it is possible to controllably align or rotate disc shaped
cells\textemdash chloroplasts of {\it Spinacia oleracea}\textemdash%
in a plane polarised Gaussian beam trap, using optical torques
resulting predominantly from circular polarisation induced in the
transmitted beam by the non-spherical shape of the cells.

\end{abstract}

\section{Introduction}

Optical tweezers are widely used for the manipulation of cells and their
internal structures~\cite{ashkin1997}.
A strongly focussed laser beam is used to
apply piconewton forces, which is sufficient to trap or to move cells
in three dimensions. However, there is poor control of the orientation
of cells within the trap. While a variety of methods to rotate
microscopic objects have been demonstrated, they are either
dependent on
absorption~\cite{friese1996,friese1998ol},
making them unsuitable for biological
applications, restricted to special types of
particles~\cite{higurashi1997,friese1998nature,galajda2002}, or
make use of multiple beams or special types of
beams~\cite{paterson2001,oneil2002,santamato2002}. The latter
methods can, in principle, be used for the rotation of biological
specimens, but either add a great deal of complexity to the required
apparatus, or use expanded non-symmetric focal spots, reducing
the possibility of three-dimensional trapping.

The alignment of elongated dielectric particles within static fields suggests
that it might be possible to align non-spherical trapped objects simply
by using a plane polarised trapping beam.
Unlike rod shaped objects, which tend to align along the beam
axis~\cite{ashkin1987}
leaving no remaining asymmetry with respect to the plane of polarisation,
disc shaped objects, aligned with the beam axis, retain an asymmetric
shape perpendicular to the beam axis and should also align with
the plane of polarisation.
Such an alignment scheme has the dual advantages
of being easy to implement using an existing optical tweezers apparatus,
and providing maximum trapping efficiency. The torques acting to
align a dielectric object within a static field result from the induced
polarisation within the object being at an angle to the applied field.
It would, however, be questionable to na\"{\i}vely apply the static
field results to trapping by optical fields.

We demonstrate experimentally that sufficient torque is generated for
the technique to be useful for biological
specimens\textemdash chloroplasts of {\it Spinacia oleracea}
in our experiments\textemdash%
and calculate the optical force and torque
using a full electromagnetic wave solution of Maxwell's equation to
show unambiguously that the torque can be accounted for by
the non-spherical shape of the chloroplasts.

\section{Observation of alignment of chloroplasts}

A standard optical tweezers setup using an inverted microscope was used.
The only modification required for the alignment experiments was the
addition of a half-wave plate on a rotatable mount
to control the direction of the plane of
polarisation (see figure 1). A Nd--YAG laser at
1064\,nm was used, focussed by a 100$\times$ oil-immersion object of
numerical aperture 1.3 to produce the trap.
 
\begin{figure}[htb]
\centerline{\includegraphics[width=0.7\textwidth]{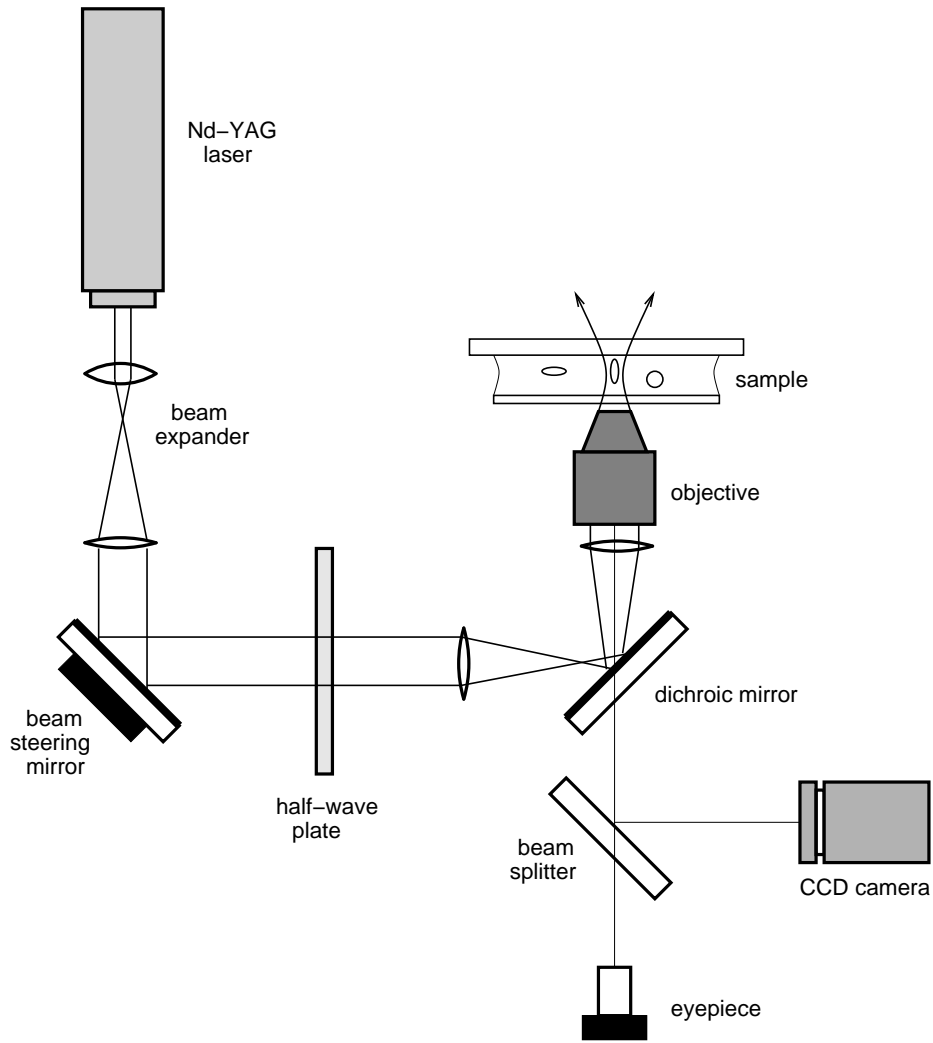}}
\caption{Schematic diagram of optical tweezers system for alignment
of biological cells. The apparatus is a conventional optical tweezers
setup with the addition of a half-wave plate in a rotatable mounting,
allowing the plane of polarisation of the initially plane polarised
trapping beam to be rotated.}
\end{figure}

Chloroplasts were isolated from fresh spinach leaves
(\textit{Spinacia oleracea} L.) using a buffer solution containing
0.4\,M sucrose, 0.05\,M HEPES, 0.01\,M KCl, 0.0001\,M MgCl$_2$ with
a pH of 7.8.
Chloroplasts were three-dimensionally trapped in the
buffer solution using a beam power of 30\,mW at the specimen plane.
Trapped chloroplasts were observed to align with the plane of polarisation
of the trapping beam. As the plane of polarisation of the trapping beam
was rotated, the chloroplasts rotated so as to re-align with the beam, taking
on the order of a second or two to do so. Alignment of a chloroplast is
shown in figure 2.

\begin{figure}[htb]
\centerline{
\includegraphics[width=0.23\textwidth]{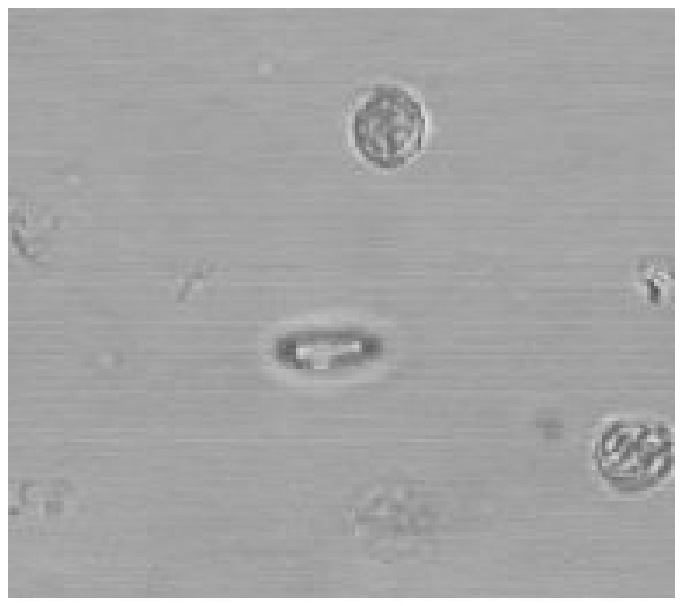}
\includegraphics[width=0.23\textwidth]{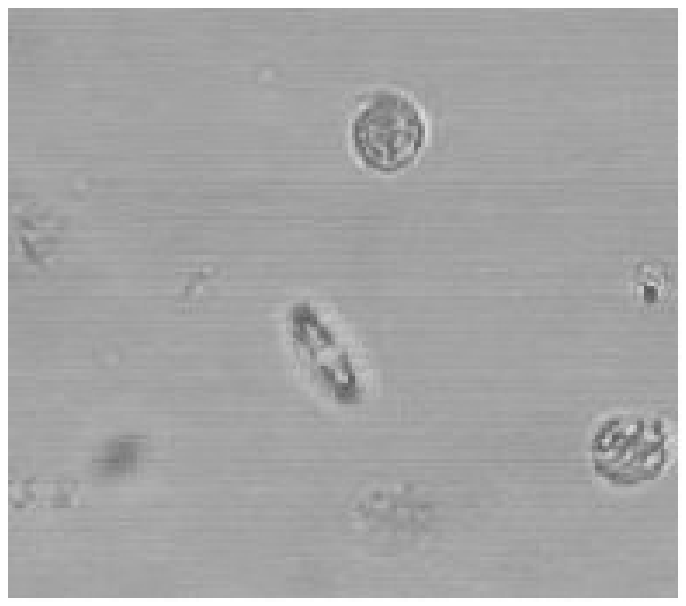}
\includegraphics[width=0.23\textwidth]{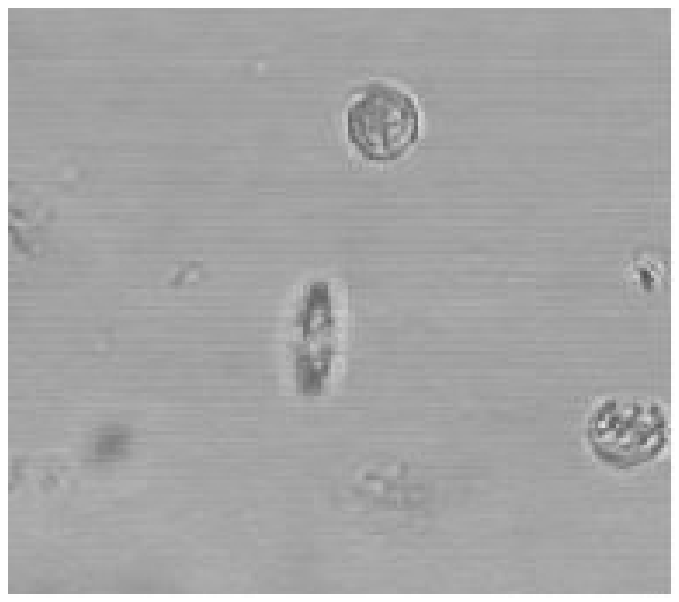}
\includegraphics[width=0.23\textwidth]{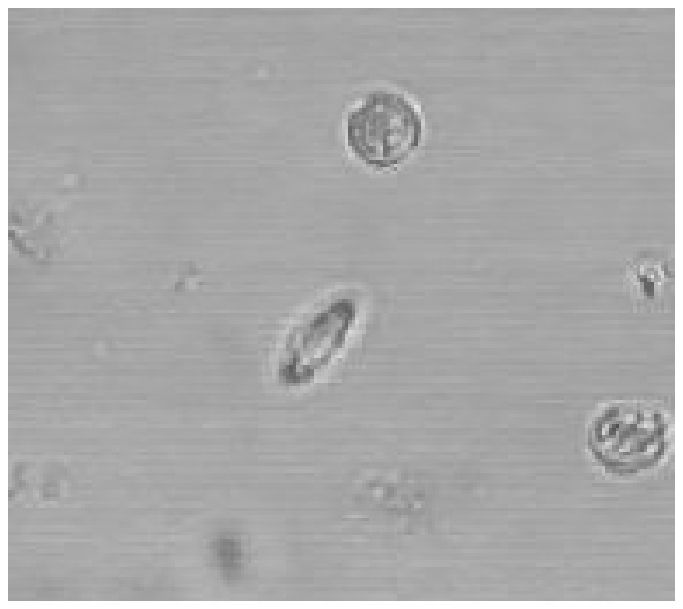}
}
\caption{Alignment of chloroplasts with the plane of polarisation
of the trapping beam. The frames show rotation of a chloroplast
as the plane of polarisation of the trapping beam is rotated. In each
frame, the chloroplast lies in the plane of polarisation. The chloroplast
is approximately 4\,$\mu$m long along the long axis.}
\end{figure}

Since the observed alignment is similar to that expected from birefringent
particles~\cite{friese1998nature},
it necessary to exclude this possibility. This was done by replacing
the half-wave plate with a quarter wave plate to produce a circularly
polarised trapping beam. If the chloroplasts were sufficiently
birefringent so as to align with the plane polarised beam, the circularly
polarised beam should have caused them to rotate at a constant
rate~\cite{friese1998nature}.
Chloroplasts were not observed to rotate or
align in the circularly polarised beam. No birefringence could be
detected when viewing chloroplasts between crossed polarisers.

\section{Optical force and torque}

Optical forces and torque are a necessary result of the conservation
of momentum and angular momentum when a particle scatters light, changing
the momentum or angular momentum. Therefore, if the scattering of the
trapping beam by the trapped cell can be calculated, the optical force
and torque can be calculated~\cite{nieminen2001jqsrt}.

The chloroplasts that we wish to model scattering by are
too large for the Rayleigh (ie small particle) approximation to be
valid, and too thin for the geometric optics approximation to be valid.
Therefore, a full EM wave scattering calculation is needed. We calculate
the scattered fields using the \textit{T}-matrix
method~\cite{mishchenko_book,waterman1971,mishchenko1991,nieminen2001cpcb},
in which,
as in generalised Lorenz-Mie theory (GLMT)~\cite{ren1996},
the incident and scattered fields
are expanded in terms of vector spherical wavefunctions (VSWFs), which
are the electric and magnetic multipole fields. Unlike GLMT, the
\textit{T}-matrix method does not require the surface of the particle
to be a constant surface in a separable coordinate system. 

Apart from being computationally well-suited for optical tweezers
calculations since the \textit{T}-matrix for a given particle
needs to be calculated only
once~\cite{nieminen2001jqsrt,nieminen2001cpcb},
the mathematical formulation of
the \textit{T}-matrix method is physically enlightening since the VSWFs
are simultaneous eigenfunctions of the total angular momentum operator,
with eigenvalues $[n(n+1)]^{1/2}$, and the $z$-component of angular
momentum operator, with eigenvalues $m$.

In our \textit{T}-matrix calculations, the incoming and outgoing
fields are expanded in terms of incoming and outgoing VSWFs:
\begin{eqnarray}
\mathbf{E}_\mathrm{inc}(\mathrm{r}) & = & \sum_{n=1}^\infty \sum_{m = -n}^n
a_{nm} \mathbf{M}_{nm}^{(2)}(k\mathrm{r}) +
b_{nm} \mathbf{N}_{nm}^{(2)}(k\mathrm{r}).
\label{incoming_expansion} \\
\mathbf{E}_\mathrm{scat}(\mathrm{r}) & = & \sum_{n=1}^\infty \sum_{m = -n}^n
p_{nm} \mathbf{M}_{nm}^{(1)}(k\mathrm{r}) +
q_{nm} \mathbf{N}_{nm}^{(1)}(k\mathrm{r}).
\label{outgoing_expansion}
\end{eqnarray}
where the VSWFs are
\begin{eqnarray}
\mathbf{M}_{nm}^{(1,2)}(k\mathrm{r}) & = & N_n h_n^{(1,2)}(kr)
\mathbf{C}_{nm}(\theta,\phi) \\
\mathbf{N}_{nm}^{(1,2)}(k\mathrm{r}) & = & \frac{h_n^{(1,2)}(kr)}{krN_n}
\mathbf{P}_{nm}(\theta,\phi) + \nonumber \\
& & N_n \left( h_{n-1}^{(1,2)}(kr) -
\frac{n h_n^{(1,2)}(kr)}{kr} \right) \mathbf{B}_{nm}(\theta,\phi)
\end{eqnarray}
where $h_n^{(1,2)}(kr)$ are spherical Hankel functions of the first and second
kind, 
$N_n = [n(n+1)]^{-1/2}$ are normalisation constants, and
$\mathbf{B}_{nm}(\theta,\phi) = \mathbf{r} \nabla Y_n^m(\theta,\phi)$,
$\mathbf{C}_{nm}(\theta,\phi) = \nabla \times \left( \mathbf{r}
Y_n^m(\theta,\phi) \right)$, and
$\mathbf{P}_{nm}(\theta,\phi) = \hat{\mathbf{r}} Y_n^m(\theta,\phi)$
are the vector spherical
harmonics~\cite{mishchenko_book,waterman1971,mishchenko1991,jackson},
and $Y_n^m(\theta,\phi)$ are normalised scalar spherical harmonics. The usual
polar spherical coordinates are used, where $\theta$ is the co-latitude
measured
from the $+z$ axis, and $\phi$ is the azimuth, measured from the $+x$ axis
towards the $+y$ axis. It should be noted that our division of the fields
into a purely incoming incident field and an outgoing scattered field is
unusual; it is much more common to include the outgoing field resulting
from the incident field in the absence of a scatterer as part of the incident
field. Both formulations are equivalent~\cite{nieminen_tmatrix}; our choice
simplifies the expressions for optical force and torque. In practice,
the field expansions and the \textit{T}-matrix must be terminated
at some finite $n = N_{\mathrm{max}}$ chosen so that the numerical results
converge with sufficient accuracy~\cite{nieminen_tmatrix,nieminen_focussed}.

The normalised torque about the $z$-axis acting on the trapped particle is
\begin{equation}
\tau_z = \sum_{n=1}^\infty \sum_{m = -n}^n m ( |a_{nm}|^2 + |b_{nm}|^2
- |p_{nm}|^2 - |q_{nm}|^2 ) / P
\end{equation} 
in units of $\hbar$ per photon, where
$ P = \sum_{n=1}^\infty \sum_{m = -n}^n |a_{nm}|^2 + |b_{nm}|^2 $
is proportional to the incident power (omitting a unit conversion
factor which will depend on whether SI, Gaussian, or other units
are used).
This torque includes contributions from both spin and orbital components;
the spin torque about the $z$-axis is given by~\cite{crichton2000}
\begin{eqnarray}
\sigma_z & = & \frac{1}{P} \sum_{n=1}^\infty \sum_{m = -n}^n
\frac{m}{n(n+1)} ( |a_{nm}|^2 + |b_{nm}|^2 - |p_{nm}|^2 - |q_{nm}|^2)
\nonumber \\ & & - \frac{2}{n+1}
\left[ \frac{n(n+2)(n-m+1)(n+m+1)}{(2n+1)(2n+3)} \right]^{\frac{1}{2}}
\nonumber \\ & & \times
\mathrm{Im}( a_{nm} b_{n+1,m}^\star + b_{nm} a_{n+1,m}^\star
- p_{nm} q_{n+1,m}^\star - q_{nm} p_{n+1,m}^\star ).
\end{eqnarray}
The remainder of the torque is the orbital contribution.
The axial trapping efficiency $Q$ is~\cite{crichton2000}
\begin{eqnarray}
Q & = & \frac{2}{P} \sum_{n=1}^\infty \sum_{m = -n}^n
\frac{m}{n(n+1)} \mathrm{Re}( a_{nm}^\star b_{nm} - p_{nm}^\star q_{nm} )
\nonumber \\ & & - \frac{1}{n+1}
\left[ \frac{n(n+2)(n-m+1)(n+m+1)}{(2n+1)(2n+3)} \right]^{\frac{1}{2}}
\nonumber \\ & & \times
\mathrm{Im}( a_{nm} a_{n+1,m}^\star + b_{nm} b_{n+1,m}^\star
- p_{nm} p_{n+1,m}^\star - q_{nm} q_{n+1,m}^\star ).
\end{eqnarray}

The expansion coefficients of the outgoing (ie scattered) field are
found from the expansion coefficients of the incoming field using the
\textit{T}-matrix:
\begin{equation}
\mathbf{p} = \mathbf{T} \mathbf{a}.
\end{equation}
where $\mathbf{a}$ and $\mathbf{p}$ are vectors formed from the
expansion coefficients of the incident wave ($a_{nm}$ and $b_{nm}$) and
the scattered wave ($p_{nm}$ and $q_{nm}$).
The expansion coefficients of the incoming field are calculated using
far-field point-matching~\cite{nieminen_focussed}, and the \textit{T}-matrix
is calculated using a row-by-row point-matching method, exploiting
symmetry of the particle when possible~\cite{nieminen_tmatrix}. 

\section{Optical force and torque on trapped chloroplasts}

We model chloroplasts as oblate spheroids. Our numerical results are
calculated for a spheroid of aspect ratio 4, with semi-minor axis
of 0.5\,$\mu$m and semi-major axes 2.0\,$\mu$m.
This is typical of the size
and aspect ratio of the chloroplasts aligned in our experiment.
In reality, one face of the chloroplasts is concave~\cite{bayoudh2001}, but our
simplified geometry serves as a general model for disc shaped cells
and organelles, and adequately models the almost disc shaped
chloroplasts.
At the trapping
wavelength of 1064\,nm, the refractive index of the chloroplast is
assumed to be 1.4, which is typical of organelles such as mitochondria and
nuclei as well as chloroplasts. Therefore, the results for chloroplasts,
including both experimental observations and theoretical caclulations,
are also applicable to other organelles of similar size and shape.
The refractive index of the sucrose solution in which the chloroplasts are
trapped is 1.35. The trapping beam was chosen to reproduce the
focal spot size in the trap. The focal spot is elliptical, with
radii of 0.60\,$\mu$m and 0.54\,$\mu$m along and perpendicular
to the plane of polarisation,  respectively. The ellipticity
of the beam spot is a consequence of strongly focussing
a plane polarised vector beam~\cite{nieminen_focussed,sales1998}.

The axial trapping efficiency is shown in figure 3. The equilibrium position
of the chloroplast in the observed orientation (aligned
with the plane of polarisation) is with its centre 1.20\,$\mu$m
beyond the beam focus.
Since the chloroplasts are three-dimensionally trapped during the course
of the experiment, torques are evaluated when the chloroplast is at this
equilibrium position within the trap.

\begin{figure}[htb]
\centerline{\includegraphics[width=0.7\textwidth]{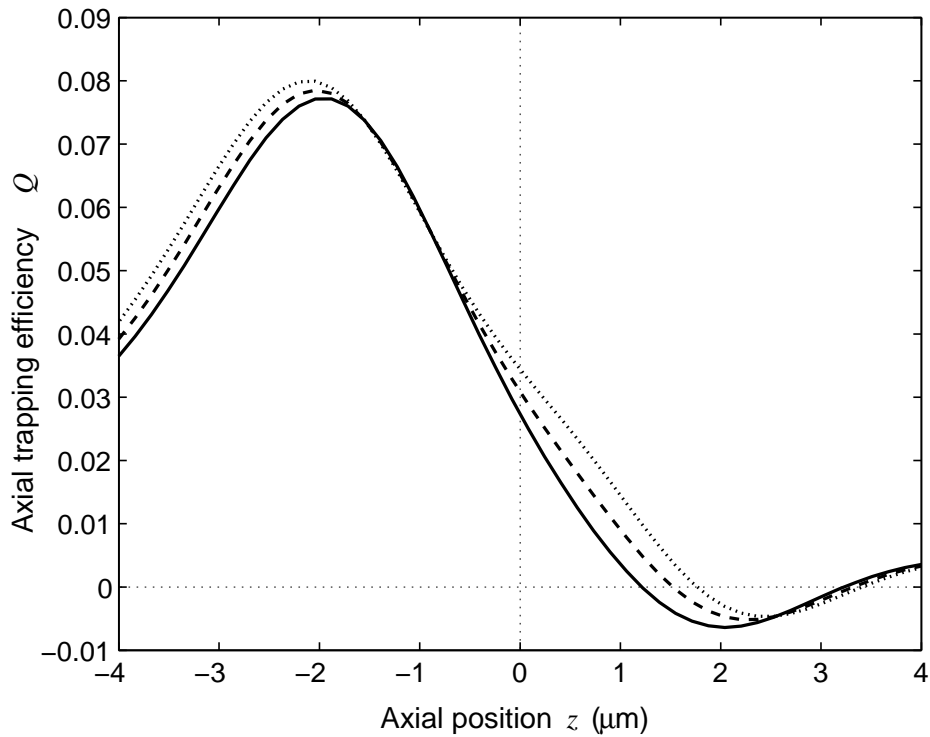}}
\caption{Axial trapping efficiency. The trapping efficiency is shown
for an oblate spheroidal chloroplast with semi-minor axis 0.5\,$\mu$m
and semi-major axes 2\,$\mu$m. The trapping efficiency is shown for
chloroplasts aligned along the beam axis, at varying angles to the
plane of polarisation. The force is shown for chloroplasts aligned with the
plane of polarisation (solid line), at 45$^\circ$ to (dashed line), and
perpendicular to the plane of polarisation (dotted line). The
equilibrium position in all cases is with the centre of the chloroplast
slightly beyond the focal plane of the beam located at $z = 0$.}
\end{figure}

The torques acting to align the plane of the chloroplast with the beam
axis and the plane of polarisation are shown in figure 4. The torque
acting to align the chloroplast with the beam axis depends on the axis
about which the chloroplast rotates. However, this torque is always
much larger than the torque aligning the chloroplast with the plane of
polarisation. Noting that the drag torque acting on the chloroplast
will be on the order of the drag torque that would act on a sphere of
radius equal to the semi-major axis of the chloroplast, we
can estimate that
the drag torque will be of order
$\tau_{\mathrm{drag}} = 8\pi \mu a^3 \Omega = D \Omega$
where $\mu$ is the dynamic viscosity of the medium
($\approx 10^{-3}$\,Nsm$^{-2}$ for water, and
$\approx 1.48 \times 10^{-3}$\,Nsm$^{-2}$ for the sucrose solution in our
experiments), $a$ is the semi-major axis of
the chloroplast, and $\Omega$ is the rotation rate of the chloroplast.
For the chloroplast considered here, the drag coefficient
$D \approx 0.3$\,pN$\mu$ms/rad. For the beam power used in our experiments,
30\,mW, the mean torque acting to align the chloroplast with the beam axis
is about 3.8\,pN$\mu$m, giving a mean rotation speed of
$\approx 13$\,rad/s, so the chloroplasts can be expected to take on the
order of a tenth of a second,
or less depending on the initial position, to align with the beam
axis. The torque acting to align the chloroplast with the
plane of polarisation is much smaller, with a mean value of
0.22\,pN$\mu$m, giving a mean rotation speed of $\approx 0.73$\,rad/s,
and an alignment time on the order of 2\,s. This is consistent with
observations made during the experiment.

\begin{figure}[htb]
\centerline{\includegraphics[width=0.7\textwidth]{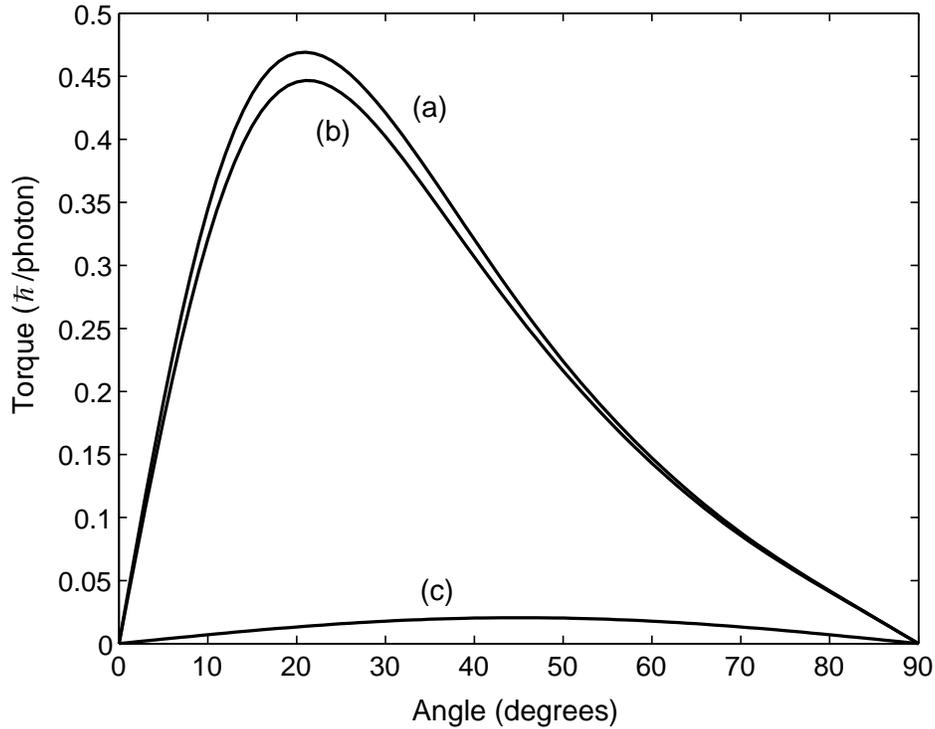}}
\caption{Alignment torque on chloroplasts. The torque is shown
for alignment with the beam axis with the axis of rotation
(a) perpendicular to the plane of polarisation and (b) parallel
to the plane of polarisation, and (c) for alignment with the plane
of polarisation by rotation about the beam axis after the chloroplast
has aligned with the beam axis. The angle is the angle between the plane
of the chloroplast and the beam axis/plane of polarisation.}
\end{figure}

The numerical results agree well with the observed behaviour of the
chloroplasts, allowing us to unambiguously identify the non-spherical
shape of the chloroplasts as the cause of their alignment with the
plane of polarisation of the beam. The spin and orbital contributions
to the alignment torque are shown in figure 5.
The spin torque is approximately 23 times larger than the orbital torque,
for all angles between the plane of the chloroplast and the plane of
polarisation for which the torque is non-zero. Therefore, the torque
is predominantly due to circular polarisation of the transmitted
light resulting from scattering by the chloroplast.

\begin{figure}[htb]
\centerline{\includegraphics[width=0.7\textwidth]{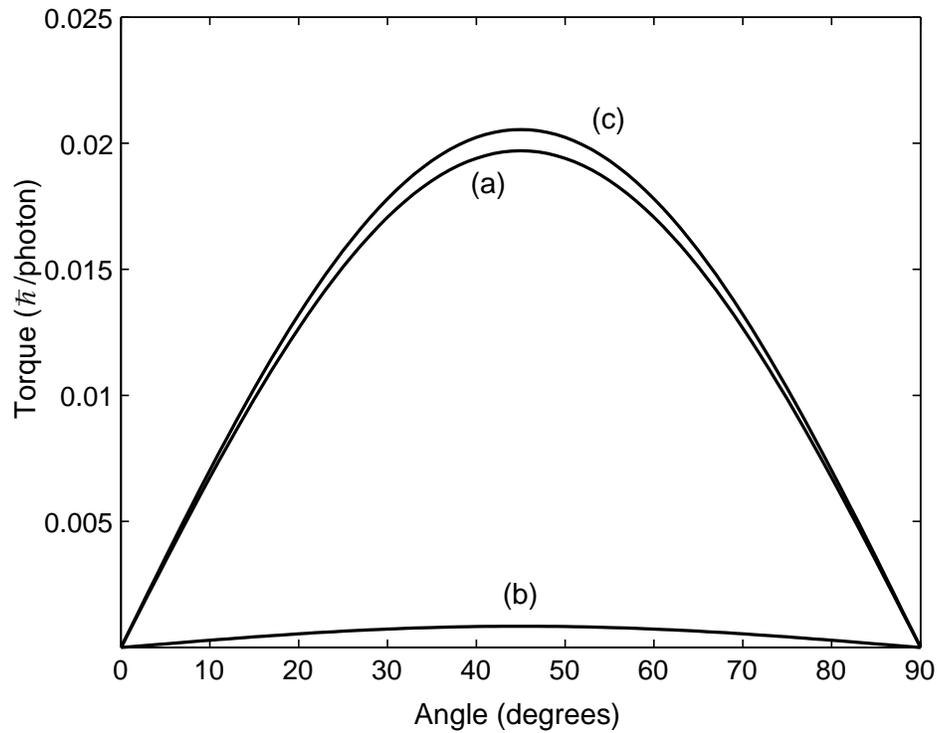}}
\caption{Spin and orbital components of torque. The (a) spin and
(b) orbital components of (c) the total torque acting to
align a chloroplast with the plane of polarisation of the trapping beam
are shown. The chloroplast is already aligned with the beam axis, and the
torque is about the beam axis.}
\end{figure}

The polarisation of the transmitted light is affected by the scattering
process due to the differing dielectric polarisability of the chloroplast
parallel to its long and short axes~\cite{jones1945}. This results in a
phase difference between the plane-polarised components in the transmitted
beam parallel to the long and short axes of the chloroplast, which generally
produces an elliptically polarised transmitted beam. This is similar to the
action of a birefringent particle, and this effect is termed
\emph{form birefringence} or \emph{shape birefringence}~\cite{born1980}.
The small
orbital torque is due to the interaction between the elliptical
focal spot of the beam and the shape of the chloroplast.
The ellipticity of the focal spot results when a plane-polarised
beam of circular cross-section is focussed to a diffraction-limited
spot~\cite{nieminen_focussed,sales1998}.

The dependence of the torque on the size and shape of an oblate spheroid
is shown in figure 6. Since the difference in polarisability of the
spheroid along the long and short axes increases with increasing
aspect ratio, the torque increases as the aspect ratio increases.
For small particles, the torque increases as the particle size increases,
because the optical thickness and therefore the phase shift between the
plane-polarised components parallel to the long and short axes, and the
particle intercepts more of the beam as its cross-sectional area increases.
Once the particle becomes larger than the beam, the illuminated portion of
the particle is approximately rotationally symmetric about the beam axis,
and the torque becomes smaller with increasing particle size. The largest
possible torque is generated when the particle is as large as possible, while
retaining maximum rotational asymmetry of the illuminated portion of the
particle. This occurs when the semi-minor axis is approximately equal to the
beam waist radius.

\begin{figure}[htb]
\centerline{\includegraphics[width=0.7\textwidth]{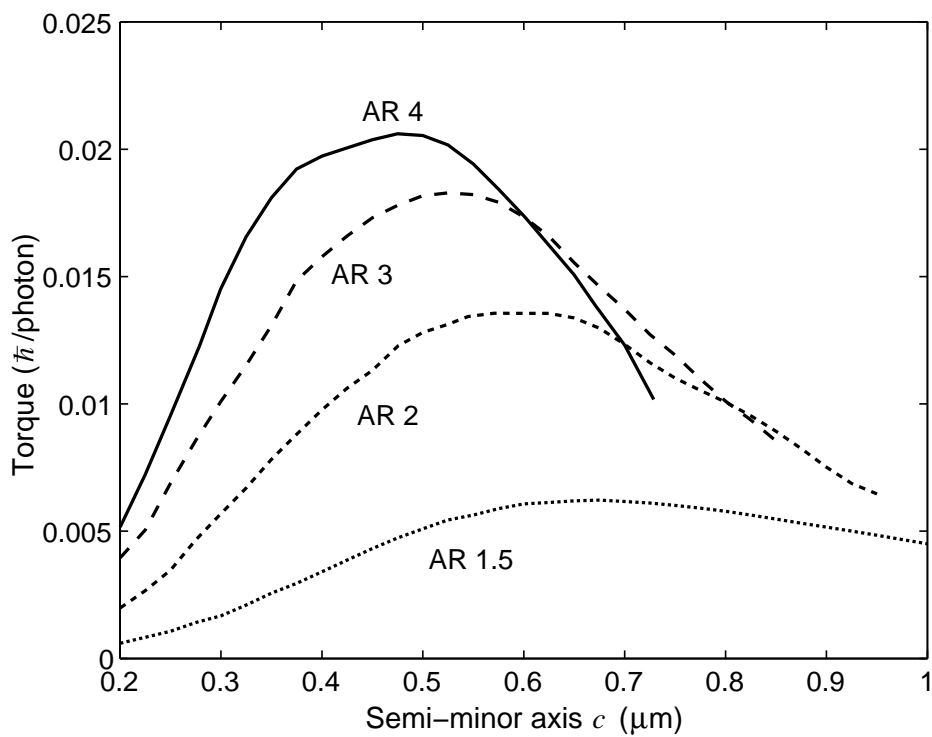}}
\caption{Dependence of torque on aspect ratio. The maximum torque
acting to align an oblate spheroid with the plane of polarisation, which
occurs when the angle between them is $45^\circ$, is shown for
spheroids with aspect ratios of 1.5, 2, 3, and 4.}
\end{figure}

Similar alignment effects can be expected for other types of non-spherical
particles in optical traps; alignment of short rods has been recently
reported~\cite{bonin2002}. Since the alignment torque due to non-spherical
shape is
relatively small compared to the torque that can be obtained by using
birefringent particles, it might not be observed in all cases;
for example, Cheng et al. recently trapped thin PMMA disks, but
failed to observe alignment~\cite{cheng2002}.

\section{Conclusion}

We have demonstrated that the non-spherical shape of chloroplasts is
sufficient to allow controllable alignment with optical tweezers
using a single plane-polarised Gaussian beam. The chloroplast
aligns with one long axis along the beam axis, and the other long axis in
the plane of polarisation of the beam. Since the plane of polarisation
of the trapping beam can be rotated using a half-wave plate, the
orientation of the chloroplast can be controlled. 
This method of alignment can be applied to other non-spherical
objects that have two long axes.
Since the diffraction-limited
focal spot of a strongly focussed plane polarised beam is 
slightly elliptical in shape,
the focal spot shape might contribute
somewhat to the torque. However, our calculations show that
the torque is predominantly due to
circular polarisation induced in the transmitted beam by the shape of
the chloroplast.

\section*{Acknowledgements}

Part of this work was supported by The University of Queensland and 
the Australian Research Council.

\end{document}